\documentstyle[twocolumn,aps,prl,epsf,psfig]{revtex}
\begin{document}

\title{A Maximum Entropy Method of Obtaining Thermodynamic Properties from
Quantum Monte Carlo Simulations}
\author{Carey Huscroft, Richard Gass, and Mark Jarrell}
\address{Physics Department, University of Cincinnati,
Cincinnati, OH 45221-0011}
\address{\mbox{ }} \address{\parbox{14cm}{\rm \mbox{ }\mbox{ }
We describe a novel method to obtain thermodynamic properties of
quantum systems using Baysian Inference -- Maximum Entropy techniques.
The method is applicable to energy values sampled at a discrete set
of temperatures from Quantum Monte Carlo Simulations.
The internal energy and the specific heat
of the system are easily obtained as are errorbars on these
quantities.  The entropy and the free energy are also obtainable.
No assumptions as to the specific functional form of the energy are made.
The use of {\it a priori} information, such as a sum
rule on the entropy, is built into the method.  As a non-trivial
example of the method, we obtain the specific heat of the
three-dimensional Periodic Anderson Model.
}}\address{\mbox{ }} \address{\mbox{ }}
\address{\parbox{14cm}{\rm \mbox{ }\mbox{ } PACS numbers:
02.50.Rj, 65.50.+m, 02.70.Lq, 71.45.-d}} \maketitle
\narrowtext

\section{\it Introduction}
\label{INT_SECTION}

The problem of obtaining thermodynamic properties from Quantum
Monte Carlo (QMC) simulations is one of long--standing
interest.\cite{LONGSTAND}.  Although the internal energy, i.e., the
expectation value of the Hamiltonian, is one of the easiest
quantities to obtain via QMC, the free energy is almost impossible
to obtain directly in a simulation.
Likewise, the specific heat, i.e.,
the temperature derivative of the internal energy, is very difficult to
obtain directly.  Hence, one must turn to indirect methods.

Several methods to obtain the thermodynamic properties of model
systems via QMC  have been proposed,
but all suffer from limitations
of one sort or another.  To a large extent, these stem from the
use of a specific functional form to fit the internal energy of the
system.  We propose a novel method to obtain the internal energy,
the specific heat, the entropy,
and the free energy as a function of temperature
via QMC which does not impose any functional form on these quantities
and alleviates several other problems in the current methods.  Our
technique relies on probability theory and Maximum Entropy to obtain
the {\it most probable} thermodynamic functions consistent with the
QMC data and prior knowledge, such as a sum rule on the system's
entropy.

In the remainder of this paper, Sec.~\ref{BAK_SECTION}
reviews currently used
techniques to obtain thermodynamic properties from QMC, their
limitations, and the desirable features of a new technique.
Sec.~\ref{OVR_SECTION} contains a brief overview of our method to obtain
thermodynamic quantities from QMC data.
In Sec.~\ref{MEM_SECTION}, we review the theoretical underpinnings
of the method, Maximum Entropy.  Sec.~\ref{ALG_SECTION}
sets forth the algorithmic
details of our method.  In order to test our method, we apply
it to a non-trivial problem -- the 3d Periodic Anderson Model -- in
Sec.~\ref{PAM_SECTION}.  Our summary is given in Sec.~\ref{SUM_SECTION}.

\section{\it Background}
\label{BAK_SECTION}

The free energy and its derivatives, including the specific heat, provide 
experimentally relevant insight into a system's temperature evolution 
and phase transitions.  Unfortunately, both direct and indirect QMC 
measurements of such quantities 
are notoriously difficult to make.
To understand why, we discuss two methods
for obtaining thermodynamic quantities in this section.

Typically, one tries to obtain the thermodynamic properties of a system
by performing QMC simulations at various discrete temperatures, then
fitting the resultant energy data to a functional form.  Generally,
this functional form is not known, so a physically-motivated form must be
chosen.  The recipe is to fit the internal energy $E(T)$ to a functional form,
which may then be differentiated explicitly to obtain the specific heat
\begin{eqnarray}
C(T) = {\partial E(T) \over \partial T}.
\label{sheat}
\end{eqnarray}
From the specific heat, the entropy $S(T)$ may be calculated by
integrating
\begin{eqnarray}
S(T) = \int^{T}_{0} dT^\prime {C(T^\prime) \over T^\prime}.
\label{entropy}
\end{eqnarray}
Then, the free energy $F$ may be obtained from the relation
\begin{eqnarray}
F(T) = E(T) - T S(T).
\label{fenergy}
\end{eqnarray}

While apparently sound in principle, this prescription can 
manifest several serious problems.  For one, the derivative in Eq.~1
enhances the statistical uncertainty in the fit.
At low temperatures the procedure is
further complicated by the division by $T$ in Eq.~2.  Similar
problems emerge at low $T$ when the specific heat is evaluated directly 
$C= \left(\left<E^2\right>-\left<E\right>^2\right)/T^2$.  Thus,
there is no guarantee that the total entropy obtained by this method
\begin{eqnarray}
S_{\infty} = \int^{\infty}_{0} dT {C(T) \over T}
\label{sinfinity}
\end{eqnarray}
will equal the total infinite-temperature entropy of the system.
If not, then both the specific heat $C(T)$ and the free-energy
$F(T)$ may be unreliable.

One technique using this prescription is to fit QMC internal energy data to
a pair of functional forms, one for low-temperature data and another
for high-temperature data.\cite{DUFFY}
This method has been successfully applied, but it does have
several important drawbacks.  First,
separate functional forms in the nature of polynomials
are used for the low- and the high-temperature
regions.  These are chosen after viewing the data, based
on the shape of the $E(T)$ curve.  Presumably, care must be taken
to avoid spurious features in the derivative $C(T)$ of the internal
energy $E(T)$ at the point where the two polynomials are joined.
While an analytic function such as a polynomial expansion is a
reasonable choice for an internal energy function on a
finite-dimensional lattice, there is no guarantee that sufficient
terms have been chosen for the polynomial and the most reliable
test is a goodness of fit.  Consequently, this technique requires
a great deal of costly QMC data, especially in the neighborhood
of a phase transition, in order to insure that a reasonable goodness
of fit is obtained.

Another recently-proposed method to obtain thermodynamic properties
from QMC data is to fit the internal energy $E(T)$ to a
physically-motivated functional form.\cite{JCAMD,PAMPRL}
In this method, appropriate for a lattice simulation, the internal
energy is fit to a sum of exponentials
\begin{eqnarray}
E(T) = E_0 + \sum_n c_n e^{- n \Delta / T}.
\label{andysheat}
\end{eqnarray}
The specific heat
and the entropy are then obtained according to Eqs.~\ref{sheat} \&
\ref{entropy}, respectively.

One way to view the exponential functional form of this method is
to note that physically, one may expect different energy scales
to become important as the temperature $T$ is varied.  At those
energy scales, contributions to the internal energy $E(T)$
are effectively switched on.  This is manifest in
the fitting parameter $\Delta$; at each temperature
corresponding to the energy scale
$n \Delta$ for each of the $n$ terms in the expansion, another
term in the expansion contributes to the energy.  The amplitude of
this contribution is set by the related coefficient $c_n$.

This technique has been successfully applied.
It has at least one major advantages over the
polynomial method in that 
it does not splice together
two functions for different temperature regimes.
Nevertheless, it also relies on a goodness of fit
test to determine whether a reasonable number of fitting
parameters have been chosen.
Since it uses a gapped form for the internal
energy, it is not suitable for systems in the thermodynamic
limit, as may be studied using dynamical mean field techniques.
Furthermore, both the polynomial and the exponential fitting
schemes are inaccurate when the number of fitting parameters
is small and become ill posed as the number of fitting 
parameters become large.  Thus it is difficult to determine
how many coefficients to use.

We now present a method that overcomes these drawbacks and
possible pitfalls.  Our method incorporates additional {\em{a priori}}
information that the energy $E(T)$ increases monotonically
with the temperature $T$ so that the specific heat is positive definite, 
$C(T) \ge 0$, and in the form of the infinite temperature entropy
$S_\infty$. Thus it requires less QMC data than either of the
functional fitting methods.  It performs a search for the most probable 
energy $E(T)$ given the data and prior information, and therefore removes 
the question of determining the number of fitting parameters appearing 
when using the other methods.  It is applicable to both
finite-dimensional lattice QMC as well as QMC from non-gapped
systems.

\section{\it Overview of the New Technique}
\label{OVR_SECTION}

We start with the observation that given the appropriate
distribution function $K(\beta, \omega)$ relating the
energy $\omega$ to
the inverse temperature $\beta = 1 / T$, one can write the
internal energy as a weighted integral with
a positive-definite weight $\rho(\omega)$
\begin{eqnarray}
E(T) = -\int_{-\infty}^{\infty} d \omega \omega K(\beta, \omega) \rho(\omega) .
\label{eassume}
\end{eqnarray}
The specific heat is obtained by differentiating
\begin{eqnarray}
C(T) = {{\partial E(T)} \over {\partial T}} = -\int_{-\infty}^{\infty} d
\omega \omega {{\partial K(\beta, \omega)} \over {\partial T}} \rho(\omega) .
\label{ctassume}
\end{eqnarray}
One may then use Eqs.~\ref{entropy} and \ref{fenergy}
to obtain the entropy and the free energy, respectively.

The entire problem
of obtaining the thermodynamic properties of the system then
reduces to that of numerically inverting the integral equation
Eq.~\ref{eassume} for the weight $\rho(\omega)$ given noisy
QMC data for the internal energy $E(T)$.
This is a well-known problem for which there exists a
well-developed, powerful technique, the Maximum Entropy Method
(MEM).\cite{JARRELLandGUB}  (The MEM is discussed in the next section.)
The kernel $K(\beta, \omega)$
corresponds to a blurring function which acts on a spectrum
$\rho(\omega)$.  Typical blurring functions for a
quantum system are the Bose-Einstein
and the Fermi distribution functions, as discussed in detail in
Sec.~\ref{MEM_SECTION}.\cite{NO_CLASSICAL}

To understand the fundamental difference between other methods
and our MEM technique, it is important to understand the
questions which the two methods answer.  The other methods rely
on fitting noisy data to a functional form.  They start with 
a physically-motivated functional form for $E(T)$ and seek to find
the most likely curve of the infinitely many curves which optimize some
likelihood function such as $\chi^2$.  In reality, the functional
form of the energy is not known.  What is known is that finite
temperatures spread the excitation spectrum of a quantum system.
The question one might like to ask, ``what is the curve that fits the 
data'' is therefore ill-defined.  Without additional regularization,
an infinite number of curves fit the data and, unless the precise functional
form of the internal energy $E(T)$ is known, there is no precise
answer to this question.  However, if we know the form of the
thermal blurring function $K(\beta, \omega)$, we may
ask the question, ``given the blurring function and any other
relevant prior information that we know, what is the most probable
spectrum $\rho(\omega)$ from which this energy data $E(T)$ might
arise?''  As discussed in detail below, this is the precise
question that the MEM answers.

In addition to relying on fundamental properties of quantum
systems instead of a functional form, further
benefits also accrue from employing our MEM technique.
For example, since the MEM gives errorbars on
integrated quantities, the uncertainties in both $E(T)$
and $C(T)$ are known when obtained via our MEM technique.
Other advantages of our technique will be discussed below.
Before discussing algorithmic details, it is useful
and instructive to briefly review the MEM.

\section{\it The Maximum Entropy Method}
\label{MEM_SECTION}

The Maximum Entropy Method (MEM) is discussed in detail
elsewhere.\cite{JARRELLandGUB,BRYAN,SKILLING,JARandGUBandSILVER}
Here, we wish to review only as much of the MEM as is necessary
to understand the new technique.

The MEM is frequently used to analytically continue QMC
imaginary-time Green Function data to real frequencies.\cite{JARRELLandGUB}
However, it is a general technique that is not limited to analytic
continuation or to QMC-related problems.  In fact, the
MEM has a relatively long
history as an image reconstruction technique in photography and
dynamic light scattering problems.\cite{PRESS,BRYAN}

In such problems, the observed image is the result of Gaussian
blurring of light transmitted from a source through a medium,
such as the atmosphere.  Hence, the functional form of the
image is not known and the question of whether the observed
data fits a specific functional form is ill-defined -- there
are an infinite number of curves that fit the data!  Instead,
the best that one can do is to seek the {\it most probable}
image given the data.  This is exactly what the MEM sets out
to accomplish.

This is done using Bayesian statistics.  If there are two events,
$a$ and $b$, then by Bayes' theorem, the joint probability of
these two events is
\begin{eqnarray}
P(a,b) = P(a|b) P(b) = P(b|a) P(a),
\label{jointprob}
\end{eqnarray}
where $P(a|b)$ is the conditional probability of $a$ given $b$.
The probabilities are normalized so that
\begin{eqnarray}
P(a) = \int db P(a,b) \hspace{5pt} {\rm and} \int da P(a) = 1.
\label{probnorm}
\end{eqnarray}

In our problem, we search for the spectrum $\rho$ which maximizes
the conditional probability of $\rho$ given the data $E$,
\begin{eqnarray}
P(\rho | E) = P(E | \rho) P(\rho) / P(E).
\label{condprob}
\end{eqnarray}
Typically, one calls $P(E | \rho)$ the likelihood function and
$P(\rho)$ the prior probability of $\rho$ (or the prior).
Since we work with one set of QMC data at a time, $P(E)$ is
a constant during this procedure and may be ignored.
The prior and the likelihood functions require
more thought, and are discussed in detail in Ref.~\cite{JARRELLandGUB},
here we present the salient results of that discussion.

If the spectrum is positive-definite, we may think of it as a
un-normalized probability density:
\begin{eqnarray}
\int_{-\infty}^{\infty} d \omega \rho(\omega) < \infty.
\label{unnormal}
\end{eqnarray}
Then by Skilling,\cite{SKILLING} the prior probability
is proportional to $\exp(\alpha S)$ where $S$ is the entropy defined
relative to some positive-definite function $m(\omega)$
\begin{eqnarray}
S = \int_{-\infty}^{\infty} d \omega \big[ \rho(\omega) - m(\omega) -
\rho(\omega) \ln(\rho(\omega) / m(\omega)) \big],
\label{shannonentropy}
\end{eqnarray}
and
\begin{eqnarray}
P(\rho) \propto P(\rho | m, \alpha) \propto \exp(\alpha S),
\label{skillingP}
\end{eqnarray}
where $m(\omega)$ is the default model since in the absence of
data $\rho = m$.  Selection of the default model for this
method is discussed in Sec.~\ref{ALG_SECTION}.  The other
unknown quantity $\alpha$ is determined during the MEM
to maximize the probability of the image $\rho$ given
the data.

The likelihood function follows from the central limit theorem.
If each of the measurements $E_{i,T}$ ($E_{T} \equiv E(T)$) of the
energy at a specific temperature $T$ is independent, then in
the limit of a large number of measurements $N_d$ to determine each
$E_T$ the distribution of the $E_T$ becomes Gaussian.  The
probability of measuring a particular $E_T$ is
\begin{eqnarray}
P(E_T) = {1 \over {\sqrt{2 \pi} \sigma_T}} \exp \big[{- {1 \over
\sigma_T^2} \big( \langle E_T \rangle - E_T \big)^2 / 2} \big],
\label{probET}
\end{eqnarray}
with an error estimate given by
\begin{eqnarray}
\sigma_T^2 = { 1 \over {N_d (N_d - 1)}} \sum_i \big( \langle E_T \rangle -
E_{i,T} \big)^2
\label{sigma2}
\end{eqnarray}
and
\begin{eqnarray}
\langle E_T \rangle = {1 \over N_d} \sum_{i=1}^{N_d} E_{i,T}
\label{eave}
\end{eqnarray}
for the $N_d$ measurements of $E_{i,T}$ at temperature $T$.

Then the likelihood function $P(E | \rho)$ of measuring the set of $E$ for
a given image $\rho$ is
\begin{eqnarray}
P(E | \rho) \propto e^{- \chi^2 / 2}
\label{perho}
\end{eqnarray}
where
\begin{eqnarray}
\chi^2 = \sum_T^{{\rm all } T} {{\big( {E_{T} - \sum_{\omega}
\omega K(\omega,T) \rho(\omega)} \big)^2} \over \sigma_T^2}
\label{chi2}
\end{eqnarray}
and we have discretized the integral Eq.~\ref{eassume}.

We are now in a position to perform the MEM and find the most probable
image $\rho$ given the data $E_T$.  We wish to maximize the
joint probability of the image or weight $\rho$ given the data
$E$; the default model $m$; and the Lagrange multiplier $\alpha$
\begin{eqnarray}
P(\rho | E, m, \alpha) &\propto& P(E | \rho) P(\rho | m, \alpha) \nonumber \\
&=& {\exp(\alpha S - \chi^2 / 2) \over {Z_S Z_L}}
\label{BIG_PROB}
\end{eqnarray}
where $Z_S$ and $Z_L$ are normalization factors, independent of the
image.  For a fixed $\alpha$ and the given data $E$, the
most probable image $\hat{\rho}(\alpha)$ is the one that maximizes
$Q = \alpha S - \chi^2 / 2$.  This may be found, for example,
using Newton's method.

The details of implementing a MEM code and finding $\alpha$ are given
elsewhere.\cite{JARRELLandGUB}  We will not repeat that presentation
here.  A MEM code written according to Ref.~\cite{JARRELLandGUB}
is recommended for performing the technique we discuss
herein.\cite{NOREBIN}  Having discussed the general MEM
formalism, we now turn to the specific algorithmic
details of our new technique.

\section{\it Algorithmic Details of the New Technique}
\label{ALG_SECTION}

We desire to express the internal energy of the system
as an integral over a density of energy levels
times a relation between energy and temperature, according
to Eq.~\ref{eassume}.  To that end, we make the following
Ansatz
\begin{eqnarray}
E(T) = - \int_{-\infty}^{\infty} d \omega \omega \big[ F(\beta, \omega) 
\rho_F(\omega) + B(\beta, \omega) \rho_B(\omega) \big],
\label{ansatz}
\end{eqnarray}
where $F$ and $B$ are the Fermi- and Bose-distribution functions,
respectively
\begin{eqnarray}
F(\beta, \omega) = {1 \over {1 + e^{\beta \omega}}} \nonumber \\
B(\beta, \omega) = {1 \over {1 - e^{\beta \omega}}}
\label{fermibose}
\end{eqnarray}
for $\beta = 1 / T$ (we have set the Boltzmann constant equal to
unity $k_B = 1$).

This Ansatz corresponds roughly to describing the energetics
of the system as consisting of separate linear contributions from
Fermi- and Bose-excitations and imposes the constraint that
the corresponding energy $E(T)$ increases monotonically with
temperature $T$ so that $C(T) \ge 0$.
In addition, since the degeneracy of the
ground state and the total number of accessible states is generally
know, the infinite temperature entropy is generally known.  
$S_{\infty}$ may be obtained from
\begin{eqnarray}
S_{\infty} = - \int_{0}^{\infty} {dT \over T}  \int_{-\infty}^{\infty} d
\omega \omega \big[ {{\partial F} \over {\partial T}} \rho_F(\omega) +
{{\partial B} \over {\partial T}} \rho_B(\omega) \big]
\label{sinfinitytwo}
\end{eqnarray}
by noting that the temperature integral for the Fermi term
can be done analytically and since $\rho_B(\omega)$ is odd
the Bose term does not contribute to the integral.~\cite{NOBOSE}
The net result is that
\begin{eqnarray}
S_{\infty} = \ln 2 \int_{-\infty}^{\infty} d \omega \rho_F(\omega)
\label{fermiononlys}
\end{eqnarray}
Additional information such as this may be imposed by modifying the
prior or the likelihood function.  Given the similarity of
Eq.~\ref{fermiononlys} and Eq.~\ref{ansatz}, which appears as
part of the likelihood function, we choose the latter approach.
We introduce $S_{\infty}$ as an additional datum with a relative
error estimate $\sigma_S/S_\infty$ chosen to be approximately equal to the 
smallest relative error estimate of the energy data.
We discuss our reasons for choosing this
Ansatz in the appendix.

With this Ansatz, we are in a position to employ the MEM.
We write $K \rho$ from Eq.~\ref{eassume} as a linear combination
of $F \rho_F + B \rho_B$.  Eq.~\ref{chi2} for $\chi^2$ is modified
similarly.
We pick the default model in the following manner.  We note first
that it must be positive definite and integrable.  We employ
a Gaussian default model, which satisfies these criteria.

Once the default model is selected, the method
described in Ref.~\cite{JARRELLandGUB} may be applied straight
away.  To further illustrate the method, we now apply it to
a non-trivial model.

\section{\it Example: 3d Periodic Anderson Model}
\label{PAM_SECTION}

The 3d Periodic Anderson Model (PAM) is often used to investigate
f-electron systems, where electronic correlations are important
for the phenomena under study.  It is a simplified lattice model
in which the Coulomb interaction is limited in range to on-site
interactions only and then only within one of the two bands.
Nevertheless, it is a rich model in which the interplay between
delocalization (kinetic energy), Coulomb repulsion, Pauli exclusion,
temperature, and electron density give rise to a wide variety of
phenomena.

The periodic Anderson Hamiltonian is
\begin{eqnarray}
H &=& \sum_{k \sigma} \epsilon_k
d_{k\sigma}^{\dagger} d_{k\sigma}
+ \sum_{k \sigma} V_k
(d_{k\sigma}^{\dagger} f_{k\sigma}
+ f_{k\sigma}^{\dagger} d_{k\sigma}) \nonumber \\
&&+ U_{f} \sum_{i} (n_{if\uparrow} - \frac12)
(n_{if\downarrow} - \frac12) \nonumber \\
&&+ \sum_{i\sigma} \epsilon_{f} n_{if\sigma}
-\mu \sum_{i\sigma} (n_{if\sigma} +n_{id\sigma}) \, .
\label{pamham}
\end{eqnarray}
We choose a simple cubic structure for which,
\begin{eqnarray}
\epsilon_k &=& -2 t_{dd}\,[\cos{k_x a} + \cos{k_y a} + \cos{k_z a} ] \, ,
\nonumber \\
V_k &=& -2 t_{fd}\,[\cos{k_x a} + \cos{k_y a} + \cos{k_z a} ] \, ,
\label{eq:disperse1}
\end{eqnarray}
where $a$ is the lattice constant.  The dispersion of $V_k$ reflects
our choice of near--neighbor (as opposed to on--site) hybridization of
the $f$ and $d$ electrons.  With on--site hybridization, the PAM
is an insulator at half-filling, whereas with our intersite
hybridization choice the half-filled, symmetric PAM is metallic.

\vskip0.5cm
\begin{figure}[hbt]
\unitlength1cm \begin{picture}(7.0,5.00) \put(0.0,-1.5)
{\psfig{figure=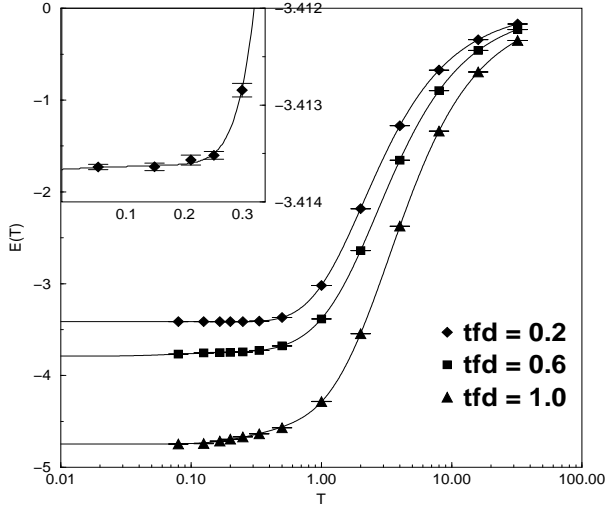,height=7.00cm,width=8.0cm,angle=-90}}
\end{picture}
\vskip1.7cm
\caption{Energy vs. temperature for various $f-d$ hybridizations $t_{fd}$.
The symbols mark QMC data, with errorbars.  The solid lines mark the
energy obtained via the MEM.  There is an excellent agreement
between the data and the MEM results throughout
the range of temperatures.  This is further illustrated in the inset,
where a magnified view of the lowest QMC temperature values for
$t_{fd}=0.2$ is shown.
}
\label{energy}
\end{figure}

The parameter values and the temperature $T$ in this work are given in
units of $t_{dd}$.  We take $U_f=6$ and explore a range of
$t_{fd}$ and $T$ values.   QMC results for this model were obtained
using the determinant algorithm,\cite{DET} which provides an exact
treatment (to within statistical errors and finite size effects) of the
correlations.  We further choose  the symmetric PAM
($\mu=\epsilon_f=0$, and thus half--filling: $\langle n_{if} \rangle =
\langle n_{id} \rangle =1$) in order to eliminate the QMC ``sign
problem,'' allowing accurate simulations at low temperatures.

This version of the PAM has been studied in this parameter regime
and is known to undergo a sharp finite-temperature
crossover at finite $t_{fd} \approx 0.6 - 0.8$
with an associated, abrupt change in the free energy which
is reflected in the specific
heat.\cite{PAMPRL,JCAMD}  These thermodynamic anomalies are
believed to be
signals of a zero-temperature metal-insulator phase transition
in the f-band which is also seen as a finite-temperature crossover
to a localized f-electron system.\cite{MAJUMDAR,HELD}
We will test our new technique by using it to reproduce the
published work.

Figure \ref{energy} shows the energy obtained via the MEM
using Eq.~\ref{ansatz} for various hybridizations
$t_{fd} = 0.2, 0.6, 1.0$ along with QMC data and errorbars
on the QMC data.  There is an excellent agreement
between the QMC data and the MEM results throughout
the range of temperatures simulated by QMC.
The quality of the agreement is further illustrated
by the inset in Fig.~\ref{energy}, which shows
a magnified view of the lowest QMC temperature values for
$t_{fd}=0.2$.

\vskip0.5cm
\begin{figure}[hbt]
\unitlength1cm \begin{picture}(7.0,5.00) \put(0.0,-1.5)
{\psfig{figure=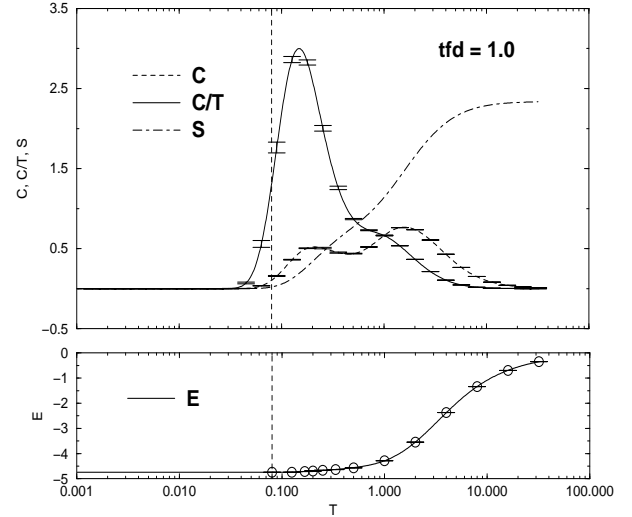,height=7.00cm,width=8.0cm,angle=-90}}
\end{picture}
\vskip1.7cm
\caption{Specific heat $C(T)$, specific heat divided by temperature $C(T)/T$,
and entropy $S(T)$ (top figure), and energy $E(T)$ (bottom figure)
as a function of temperature $T$ for a fixed hybridization $t_{fd}=1.0$.
Representative error bars are shown.  The dashed vertical line
at $T=0.08$ corresponds to the lowest QMC data point.
A peak in $C(T)/T$ appears at $T \approx 0.15$ due to singlet formation.
At a higher temperature, a hump appears in $C(T)/T$
which is believed to
be due to the suppression of charge fluctuations.
The total entropy $3.375 \ln 2$ in the system obtained by
integrating the image
from the MEM (see the text) matches both the total entropy
obtained from integrating $C(T)/T$ (Eq.~\ref{sinfinity}) and
the value of $S_{\infty} = 4 \ln 2 - S_0$ known for the
model.
}
\label{all}
\end{figure}

Once one obtains the image, the specific heat $C(T)$ is obtained
by differentiating as in Eq.~\ref{energy}.
Fig.~\ref{all} shows the specific heat $C(T)$, the
specific heat divided by the temperature $C(T)/T$,
and the entropy $S(T)$ in the top figure
as a function of temperature $T$ for a fixed hybridization $t_{fd}=1.0$.
For comparison, the bottom figure shows the energy $E(T)$.
At this hybridization, singlets are known to form at low
temperatures.~\cite{PAMPRL,JCAMD}  This is reflected in a
peak in $C(T)/T$ appearing at $T \approx 0.15$.
The singlet formation peak is also visible in $C(T)$.
A smaller peak in $C(T)$ at higher temperatures is
believed to be due to the suppression of charge fluctuations
in the $f-$ band.

The entropy may be found in two ways.  First, $S(T)$ may be
obtained by integrating $C(T)/T$ according to Eq.~\ref{entropy}.
This was done and is plotted in Fig.~\ref{all}.  The entropy
found in this manner saturates at high temperatures at
the infinite-temperature limit $S_{\infty} = 4 \ln 2 - S_0$
known for this model.~\cite{JCAMD}  Second, the infinite
temperature entropy may calculated by integrating $\rho_F$,
Eq.~\ref{fermiononlys}.   The latter estimate is
$S_{\infty} = 3.375 \ln 2$, which is also that obtained
from integrating $C(T)/T$ (Eq.~\ref{sinfinity}) and
the value of $S_{\infty} = 4 \ln 2 - S_0$ known for the
model.

\vskip0.5cm
\begin{figure}[hbt]
\unitlength1cm \begin{picture}(7.0,5.00) \put(0.0,-1.5)
{\psfig{figure=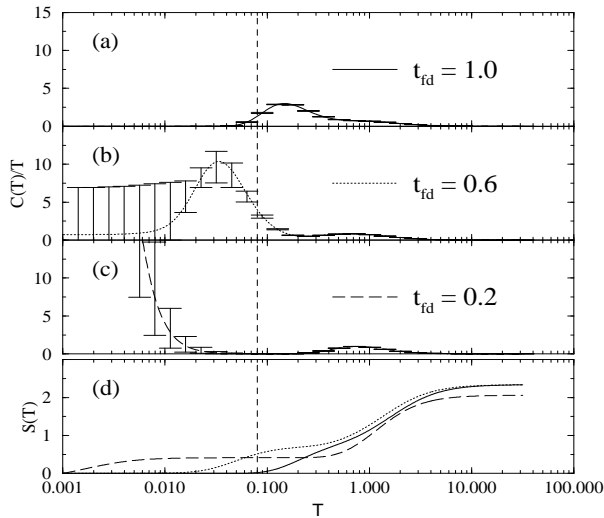,height=7.00cm,width=8.0cm,angle=0}}
\end{picture}
\vskip1.7cm
\caption{Specific heat divided by temperature $C(T)/T$ for various
$f-d$ hybridizations $t_{fd}$, (a) $t_{fd} = 1.0$, (b) $t_{fd} = 0.6$,
and (c) $t_{fd} = 0.2$.  The corresponding entropies from integrating
$C(T)/T$ (Eq.~\ref{sinfinity}) are shown in panel (d).
The dashed vertical line
at $T=0.08$ corresponds to the lowest QMC data point.
At large hybridizations, $t_{fd}=1.0$, essentially
all of the entropy in the system is quenched within the temperatures
accessed by the simulations (a).
As the hybridization decreases, antiferromagnetic
ordering at temperatures below those accessed by the simulations
becomes important and a substantial amount of entropy is not
quenched within the simulated temperatures.
The sum rule  enforces the total entropy in the
system, which appears in the results below the accessed temperatures,
yet the most probable form of the specific heat for this regime
cannot be precisely determined.This is reflected in large errorbars seen
in panels (b) and (c) for the specific heat at temperatures
below those accessed by the simulations.
}
\label{cotands}
\end{figure}

In addition to singlet formation at low temperatures for relatively
large hybridizations $t_{fd}$, the metallic PAM develops
antiferromagnetic long-range-order (AFLRO) at low temperatures
for small $t_{fd}$.  Hence, if one examines $C(T)/T$ for decreasing
hybridization $t_{fd}$, the singlet-formation peak should eventually
disappear and a low-temperature peak corresponding to AFLRO should
appear in $C(T)/T$.  Previous work has observed the disappearance
of the singlet-formation peak, but did not access a sufficiently
low temperature to observe the appearance of the AFLRO peak.~\cite{JCAMD}

Figure \ref{cotands} shows the specific heat divided by
temperature $C(T)/T$ for various
$f-d$ hybridizations $t_{fd}$, (a) $t_{fd} = 1.0$, (b) $t_{fd} = 0.6$,
and (c) $t_{fd} = 0.2$.  The corresponding entropies from integrating
$C(T)/T$ (Eq.~\ref{sinfinity}) are shown in panel (d).
Here, we observe the disappearance of the singlet peak with decreasing
hybridization $t_{fd}$.  At $t_{fd}=1.0$ there is a substantial
singlet formation peak.  At $t_{fd}=0.2$ there is no singlet formation
for any temperature accessed by the simulations, and possibly no
singlet formation even at $T=0$.~\cite{MAJUMDAR}
The intermediate hybridization $t_{fd}=0.6$ corresponds to a regime
where singlet formation occurs suddenly for low temperatures,\cite{PAMPRL}
as is reflected in the shift of the singlet peak to lower temperatures
in Fig~\ref{cotands}.

The sum rule\cite{SUMRULE}
\begin{eqnarray}
S_{\infty} = 4 \ln 2 - S_0
\label{sumruleeq}
\end{eqnarray}
for the entropy
enforces the total entropy in the system.
This is an important feature of the method and satisfying the
sum rule is one check on whether the specific heat is physically
reasonable.  However, when the system does not quench all of the
entropy by the lowest temperature accessed by the QMC simulation,
one may worry whether enforcing the sum rule will push spurious
entropy into the specific heat.  This does not happen, as shown
in Fig.~\ref{cotands}(d).  Instead, this entropy beyond that quenched
within the accessed temperatures goes below the lowest QMC
temperature, where indeed it should go on physical grounds.
However, then the extrapolation below the lowest QMC data point
is unreliable.  This unreliability of the extrapolation to
a regime where the sum rule has forced entropy below the
QMC data is seen by the large error bars on the specific heat
in this extrapolation regime.  That is, the method puts the
entropy where it belongs, but then informs one that the results
of the MEM in this regime are totally uncertain.  This is an
extremely desirable result.~\cite{LOWTFDS}

\section{\it Summary}
\label{SUM_SECTION}

We have described a novel technique to obtain the internal energy
as a function of temperature, as well as the specific heat,
the entropy, and the free energy of a system using QMC energy data
sampled at a small, finite set of temperature values.  Our
technique relies on probability theory to obtain the {\it most
probable} thermodynamic functions given the sampled QMC energy.
The question of determining the number of fitting parameters,
which plagues the other methods, is thereby removed.
An entropy sum rule or other appropriate {\it a priori} information may
also be used, if known.  The technique
was illustrated by applying it to the 3d Periodic Anderson Model.
An important benefit of the technique is that it returns not only
the thermodynamic functions, but also their uncertainties.
This is a significant improvement over the prior techniques.

\section*{\it Acknowledgements}
\label{ACK_SECTION}

We are grateful to 
J.E.~Gubernatis,
A.K.~McMahan,
R.T.~Scalettar, and
R.N.~Silver
for helpful discussions.  This work was supported by NSF grants DMR--9704021 
and DMR--9357199.  The QMC simulations were performed on the U.S.~Department
of Energy ASCI Red and Blue-Pacific computers.


\begin{appendix}
\label{ansatz_appen}
\section{The form of the energy Ansatz}
In section ~\ref{ALG_SECTION} we wrote the Ansatz for the energy as
\begin{eqnarray}
E(T) &=& 
- \int_{-\infty}^{\infty} d \omega \omega F(\beta, \omega)\rho_F(\omega)
\nonumber \\ 
&&- \int_{-\infty}^{\infty} d \omega \omega B(\beta, \omega) \rho_B(\omega)  ,
\label{Eansatz}
\end{eqnarray}
where $F$ and $B$ are the Fermi- and Bose-distribution functions,
respectively
\begin{eqnarray}
F(\beta, \omega) = {1 \over {1 + e^{\beta \omega}}} \nonumber \\
B(\beta, \omega) = {1 \over {1 - e^{\beta \omega}}}.
\label{fermiboseD}
\end{eqnarray}
Each integral in \ref{Eansatz} is a Fredholm integral equation of the first 
kind for the corresponding density with either $F$ or $B$ as the kernel.
Since these kernels are continuous, the problem of inverting Eq.~\ref{Eansatz} 
to find $E(T)$ is ill-posed \cite{Kress} and must be regularized. MEM 
provides an efficient regularization method.
In this appendix we motivate our Ansatz and discuss other
possible functional forms.

The Ansatz given in Eq. \ref{Eansatz} is not the only possible, or even
sensible choice. In general one wants to choose as an Ansatz the one 
that captures the underlying physics of the problem.  
For models such as the PAM, we 
expect a density of Fermionic excitations represented by $\rho_F$ 
associated with quasiparticle excitations.  Since quasiparticles are
conserved, their density should be positive definite which is required
when the MEM formalism is employed.  

In addition, we expect a density 
$\rho_B$ of Bosonic excitations associated 
with collective behavior such as spin waves.  With no 
Bosonic operators in the Hamiltonian, these Bosons are not conserved and 
have zero chemical potential.  Thus, $\rho_B(\omega>0) >0$ corresponding 
to the creation of such excitations, and $\rho_B(\omega<0) <0$ corresponding
to their destruction.  
This choice of $\rho_F$, $\rho_B$, and $K$ constrain the specific
heat $C(T)$ to be positive definite, $C(T) \ge 0$.
Furthermore, $\rho_B(\omega)$ is odd as required
by the Fluctuation--Dissipation theorem.  
Therefore, we may reduce the second integral in Eq.~\ref{Eansatz}
to the range $(0,\infty)$ where $\rho_B(\omega)$ is
positive semi-definite.

An Ansatz should also provide a 
faithful representation of the $E(T)$ data.  To explore this question,
we note that for a finite-sized system $E(T)$ is an analytic function
that can be expanded in a Taylor series in $T$ around $T=0$.  
Expanding $F(\beta, \omega)$ in a Sommerfeld low
temperature expansion \cite{Balescu} yields only even powers of 
$T$ in the energy. In order to get odd powers of $T$ that complete
the Taylor series, the Bosonic kernel is required.  Having the
Taylor expansion for $E(T)$, the remaining question is the
positive-definite nature of the image $\rho(\omega)$.
Even the fact that the energy is  monotonic does not
yield a mathematical constraint that $\rho_F$ and $\rho_B(\omega>0)$ are 
positive definite, yet in practice this constraint imposed by the MEM
is not a limitation.

It is possible to dispose of the Ansatz and use
a general form for the energy.
A faithful representation for systems in thermal equilibrium with
a heat bath is provided by	
\begin{eqnarray}
E(T) = - \frac{\int_{-\infty}^{\infty} d \omega \omega  \rho(\omega)
\exp(-\omega /T)}
{\int_{-\infty}^{\infty} d \omega \rho(\omega)\exp(-\omega /T)}
\label{CEansatz}
\end{eqnarray}
where $\rho(\omega)>0$ is the density of eigenstates of the 
Hamiltonian.  However, since the relationship between the density
and $E$ is non-linear, the MEM algorithm described in Sec.~\ref{MEM_SECTION}
cannot be applied without significant modification.  Furthermore, the 
representation provided by Eq.~\ref{Eansatz} in each case we have
tested has provided a fit to 
within the measured error.  Thus, the additional complications associated
with the use of Eq.~\ref{CEansatz} seem unnecessary for this
and similar systems.
However, this general technique would allow extension of the method
to Classical Monte Carlo simulations and is something we are
currently exploring.

\end{appendix}


\begin{references}
\bibitem{NO_CLASSICAL} We are
currently investigating the appropriate blurring function
for classical systems as well as the development of
a generally applicable MEM method for obtaining thermodynamic
properties from classical systems via Monte Carlo.
(See, Appendix A.)
\bibitem{NOREBIN} However, we note that the energy data $E(T)$ are
uncorrelated, so the re-binning technique
discussed in Ref.~\cite{JARRELLandGUB} does not apply here.
\bibitem{SUMRULE} The constant $S_0 = (2 N_0 / N) \ln 2$ where
$N$ is the total number of lattice sites and $N_0$ is the number
of discrete $k$ points which have $\epsilon_k = V_k = 0$ on
a finite lattice.~\cite{JCAMD}
\bibitem{LOWTFDS} The entropy $S(T)$ for $t_{fd}=0.2$ shown in
Fig.~\ref{cotands} does not saturate to the total entropy in
the system.  This is not a violation of the sum rule.  The
MEM results were not run to a sufficiently low temperature to
capture all of the entropy via Eq.~\ref{entropy}.  
When $S_{\infty}$ is obtained
by integrating the image (Eq.~\ref{fermiononlys}) the sum
rule is found to be satisfied.
\bibitem{NOBOSE} As discussed, the MEM yields the most probable image,
$\rho_{F}(\omega)$ and $\rho_{B}(\omega)$.  For the QMC
data studied, we find no finite Bose contributions to the
image, however.
\end{references}
\end{document}